\documentclass[12pt]{article}
\usepackage[style=authoryear-comp, natbib=true]{biblatex}
\bibliography{preprint.bib}
\usepackage{csquotes}
\usepackage[american]{babel}
\usepackage{authblk}
\usepackage{hyperref}
\usepackage{setspace}
\usepackage{graphicx}

\usepackage{textcomp}
\usepackage{gensymb}
\usepackage{amsmath}
\usepackage{amssymb}
\usepackage[a4paper,left=2.5cm,right=2.5cm,top=2cm,bottom=2cm]{geometry}
\usepackage{titling}

\begin{document}
\sloppy
\onehalfspacing

\title{\textsc{FLOATING FREE FROM PHYSICS:\\the metaphysics of quantum mechanics}}

\author{Raoni Wohnrath Arroyo and Jonas R. Becker Arenhart}

\affil{Graduate Program in Philosophy\\Universidade Federal de Santa Catarina \\ Florian\'opolis, Santa Catarina, Brazil \\ Research Group in Logic and Foundations of Science (CNPq)}

\date{\today}

\maketitle

\begin{abstract}
We discuss some methodological aspects of the relation between physics and metaphysics by dealing specifically with the case of non-relativistic quantum mechanics. Our main claim is that current attempts to productively integrate quantum mechanics and metaphysics are best seen as approaches of what should be called `the metaphysics of science', which is developed by applying already existing metaphysical concepts to scientific theories. We argue that, in this perspective, metaphysics must be understood as an autonomous discipline. It results that this metaphysics cannot hope to derive any kind of justification from science. Thus, one of the main motivations of such project, which is the obtaining of a scientifically respectable justification for the attribution of a single true metaphysical profile to the posits of a scientific theory, is doomed because of the emergence of metaphysical underdetermination from the outset. If metaphysics floats free from physics, which is a premise of such project of integration between these two areas, then it is always possible to attribute more than one metaphysical profile to dress physical entities.

\end{abstract}

\section{Introduction: Ontology, Metaphysics and Science} \label{sec:intro}

The relationship between metaphysics and science is nowadays less obscure than it was in the 20th century. Still, some clarification is much needed. To dive directly into the core of the debate, let us use the taxonomy provided by \citet{guayprad2017metaphysicalbox}, where \textit{metaphysics of science} seeks to enrich the scientific description of the world \textit{with} an additional metaphysical layer, and \textit{scientific metaphysics} is a project which seeks to \textit{read off} or \textit{extract} metaphysical content from science itself.

What is at stake in most of such attempts to provide for a productive relation between metaphysics and science is the degree of continuity between metaphysics and science. The main goal of such projects is to endow metaphysics with some of the epistemic credentials of current science, so that metaphysics is no longer detached from the search for the objective features of reality in which science has achieved such impressive results. All of that notwithstanding, we shall argue that the degree of continuity is very low when one is restricted to the metaphysics of science. There is always the risk that the metaphysics is, after all, floating free from science, attaining no epistemic privilege from its relation to science.

The debate is important, because it concerns the very nature, aims, and the epistemic justification of metaphysics. On the one hand, metaphysicians of science, as French, think that metaphysics can be \textit{useful} to science by providing theories that may be applied to interpret scientific concepts. A particular take on this kind of approach consists in what \citet[chap.~3]{french2014structure} called the ``Viking Approach to metaphysics'', and we shall discuss it in some detail in what follows.

On the other hand, scientific metaphysicians, or naturalists with a more radical attitude towards metaphysics, as famously advocated by \citet{Ladyman2007evmustgo}, suggest that either metaphysics is somehow more actively engaged with science itself, acquiring some of its epistemic credentials, or else it must be discontinued. Years later, \citet[p.~142]{ladyman2017apologynaturalism} clarified that the project aims the ``\textelp{} reform, not abolition'' of metaphysics as a discipline --- but that nevertheless involves the abolition of several ways of producing metaphysics, that is, those which are completely unrelated to science. The project of \textit{reading off}, or \textit{extracting} metaphysics from science we shall also call here ``radical naturalism''. The main challenge to both lines of thought is metaphysical underdetermination, the fact that more than one metaphysical theory is compatible with current scientific theories. This underdetermination plagues the realist-driven attempts to connect metaphysics with science in both approaches, and we argue that non-relativistic quantum mechanics (QM) exemplifies it nicely. We will deal with this issue in Sec.~\ref{sec:2}.

Our first result is the claim that metaphysical underdetermination makes quite unlikely the prospects of a naturalistic approach to the metaphysics of science in QM. But not all is lost. In order to see that, we need an appropriate distinction between ontology and metaphysics. Hofweber distilled such distinction in its essence:

\begin{quote}
In metaphysics we want to find out what reality is like in a general way. One part of this will be to find out what the things or the stuff are that are part of reality. Another part of metaphysics will be to find out what these things, or this stuff, are like in general ways. Ontology, on this quite standard approach to metaphysics, is the first part of this project, i.e. it is the part of metaphysics that tries to find out what things make up reality. Other parts of metaphysics build on ontology and go beyond it, but ontology is central to it, \textelp{}. Ontology is generally carried out by asking questions about what there is or what exists. \citep[p.~13]{Hofweber2016MetOnt}
\end{quote}

Such distinction, not necessarily in the same terms, can be found already in several works \citep{arenhart2012ontological, berto2015ontology, tahko2015introduction}. \citet[p.~244]{thomsonjones2017existencenature} has also put such distinction in different terms, called the ``existence question'' and the ``nature question'', which, in our terminology, is related, respectively, to \textit{ontology} and \textit{metaphysics}: ``Are there objects which are the referents of the noun phrases of that discourse \textelp{}? And if so, what sorts of things are they?''.

With this in mind, one may provide some arguments to the claim that, restricted only to ontology, a kind of naturalistic project is plausible. A first consequence is that metaphysics --- still under the distinction we shall propose --- cannot be \emph{extracted} from science, and perhaps the radical naturalization of metaphysics is threatened. The Sec.~\ref{sec:3} deals with this issue.

In Sec.~\ref{sec:4}, we deal directly with Steven French's approach to the metaphysics of science. If metaphysics is not closely related to science as naturalists expected, then how far can it go from it? Could metaphysics, as \citet{french2011metaphysical} put it, ``float free'' from science? We argue that metaphysics can and indeed does float free from science. In fact, we argue that French's theory on the relation of metaphysics and science only makes sense when metaphysics is assumed to be floating-free from physics right from the start.

French's view depends on three main tenets, briefly stated as follows. Chakravartty's Challenge: (i) that a realist understanding of science needs a metaphysical layer; Viking Approach: (ii) that already-existing metaphysical theories can and should be used to interpret scientific theories, giving content to our realism; epistemic humility: (iii) that by such interpretation our understanding of the world increases, decreasing our epistemic humility; any kind of gap between science and metaphysics is seen as unwanted and unneeded humility. Our main criticism is that if metaphysics floats free from science, then (iii) does not hold, given a kind of underdetermination of the metaphysics by the physics.

Let's go.

\section{Quantum Worldviews} \label{sec:2}

In this section, we approach what \citet[p.~193]{vanfraas1989laws} called ``the foundational question \textit{par excellence}'', which is: ``\textit{how could the world possibly be the way this theory says it is?}'', and later called ``\textit{the question of interpretation}'', which is ``[w]hat does it [a scientific theory] say the world is like?'' \citep[p.~242]{vanfrass1991quantum}. To address such questions is fundamental to understand the relation between metaphysics and science. We will do so by taking a look on how two interpretations of QM, one with collapse (QM$_{col}$) and another without collapse, but with branching (QM$_{bra}$), describe a famous quantum experiment, known as \textit{Mach-Zehnder interferometer} --- which is frequently used in the descriptions of the measurement problem. The experimental set-up consists of one monophotonic light source, two silvered mirrors ($S_2$ and $S_3$), two half-silvered mirrors or \textit{beam splitters} ($S_1$ and $S_4$), and two detectors $D_1$ and $D_2$.

One property of half-silvered mirrors is to split light beams into two. Thus, $S_1$ splits the flash $|\psi\rangle$ from the source so that the transmitted flash $|\psi_A\rangle$ goes through path $A$ and the reflected flash $|\psi_B\rangle$ goes through path $B$. The two mirrors $S_2$ and $S_3$ are then arranged in a manner to deflect $|\psi_A\rangle$ and $|\psi_B\rangle$, which then rejoin at the second half-silvered mirror in $S_4$. The path lengths of the two half beams are set equal. Given that reflected flashes undergo phase shift of $1/4$ of wavelength ($\lambda/4$), it should be expected that \textit{only} $D_1$ shows detection.

\begin{figure}
\begin{center}
\includegraphics[width=12cm]{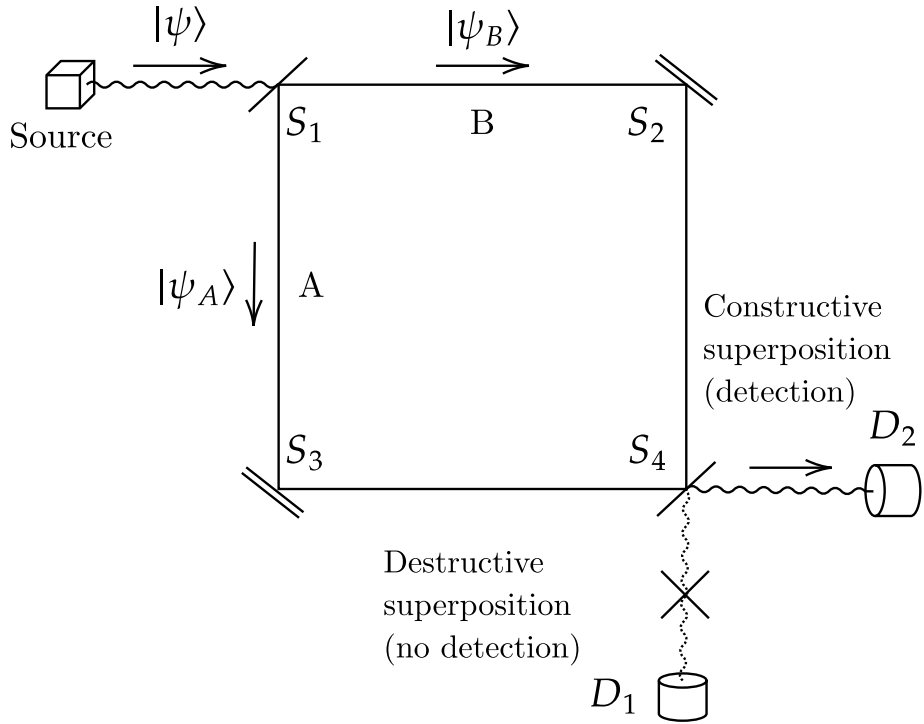}
\end{center}
\caption{Mach-Zehnder Interferometer.}
\label{fig:machzehnder}
\end{figure}

Consider the following four possible cases below concerning the trajectory of the quantum system:

\begin{enumerate}
    \item The eigenvalue corresponding to $|\psi_A\rangle$ is detected in $D_1$: at $t_1$ it is reflected by $S_1$ (phase shift $=\lambda/4$); at $t_2$ it is reflected by $S_3$ (phase shift $=2\lambda/4$); at $t_3$ reflected by $S_4$ (phase shift $=3\lambda/4$).
    \item The eigenvalue corresponding to $|\psi_B\rangle$ is detected in $D_1$: at $t_1$ it is transmitted by $S_1$ (phase shift $=0$); at $t_2$ it is reflected by $S_2$ (phase shift $=\lambda/4$); at $t_3$ it is transmitted by $S_4$ (phase shift $=\lambda/4$).
    \item The eigenvalue corresponding to $|\psi_A\rangle$ is detected in $D_2$: at $t_1$ it is reflected by $S_1$ (phase shift $=\lambda/4$); at $t_2$ it is reflected by $S_3$ (phase shift $=2\lambda/4$); at $t_3$ it is transmitted by $S_4$ (phase shift $=2\lambda/4$).
    \item The eigenvalue corresponding to $|\psi_B\rangle$ is detected in $D_2$: at $t_1$ it is transmitted by $S_1$ (phase shift $=0$); at $t_2$ it is reflected by $S_2$ (phase shift $=\lambda/4$); at $t_3$ it is reflected by $S_4$ (phase shift $=2\lambda/4$).
\end{enumerate}
 
The difference between phase shifts in $D_1$ being $\lambda/2$ results in a \textit{destructive} superposition -- hence, no detection. The same is not true in $D_2$, as $|\psi_A\rangle$ and $|\psi_B\rangle$ shows the same phase, having a \textit{constructive} superposition -- thus enabling detection. This standard setup accounts for a probability of detection in $D_2$ equal to $100\%$ and $0\%$ in $D_1$. As there is always a single photon in the experiment at once, QM yield information regarding \textit{which path} $|\psi\rangle$ followed, which is the linear combination in the form of $|\psi\rangle=\alpha|\psi_A\rangle+\beta|\psi_B\rangle$.

Suppose, however, that the half-silvered mirror $S_4$ is \textit{removed}, resulting in a case where the probability of detection is $50\%$ for $D_1$ and $50\%$ for $D_2$.

In this case, QM yields the following superposition concerning the situation of a ``\textit{which-detector}'' question, assuming that $|\psi_1\rangle$ is the eigenstate in which $|\psi\rangle$ is detected in $D_1$ and $|\psi_2\rangle$ is the eigenstate in which $|\psi\rangle$ is detected in $D_2$:

\begin{equation}
    |\psi\rangle=\tfrac{1}{\sqrt{2}}|\psi_1\rangle+\tfrac{1}{\sqrt{2}}|\psi_2\rangle \label{superpositionM-Z}
\end{equation}

The probability of measurement outcomes is given by $\textup{Prob}(|\psi_1\rangle)=|\tfrac{1}{\sqrt{2}}|^2=0,5$. The same holds for $\textup{Prob}(|\psi_2\rangle)$ as both detectors are in orthogonal paths; thus $|\psi_2\rangle=|\psi_{1}\rangle^{\bot}$ and $\langle\psi_1|\psi_{1}\rangle^{\bot}=0$. A measurement will return either $|\psi_1\rangle$ or $|\psi_2\rangle$ as an outcome, and this happens when the detectors, either $D_1$ or $D_2$ respectively, responds.

At this point, we have encountered a `smoky dragon'. \citet[p.~314]{wheeler1986smoky} once famously characterized QM as a ``\textelp{} great smoky dragon whose tail is sharply defined, whose bite is also well marked, but which in between cannot be followed''. Wheeler's analogy becomes clear as we `see' the dragon's tail as the source signal, which passes through $S_1$. Then, we `see' its head, biting the detectors $D_1$ or $D_2$. And the \textit{which-path} question brings with itself the dragon's smoky parts: we cannot be sure of what is going on in this case without the help of an interpretation, because the question itself brings the measurement problem to the surface. There is an extensive literature on many formulations of the measurement problem; for our purposes, we may state it as the \textit{which-path} question concerning the Mach-Zehnder experiment as shown in Fig. \ref{fig:machzehnder}. The job of solving the measurement problem is an interpretative one. In the following, we will see how our selected examples of interpretation account for such experiment.

Let us begin with QM$_{col}$. It says that equation \ref{superpositionM-Z} describes the whole system until the detection, when the system collapses to either $|\psi_1\rangle$ or $|\psi_2\rangle$ with equal probability. But then there's an ontological posit that enters the scene. It has been argued that the collapse is somehow related to human consciousness in \citet{wigner1961mindbody, LonBau1939theory}. So, in terms of ontology, we may say that QM$_{col}$ is ontologically committed with consciousness, hence we label such ontological aspect, provided by the interpretation, as a Consciousness-Based Interpretation (CBI). This remains at a very general level. One could go further and ask what, exactly is this consciousness in metaphysical terms. It may be entertained the idea that Consciousness \textit{Causes} the Collapse Hypothesis (CCCH) \citep{wigner1961mindbody}; it may \textit{also} be entertained the Consciousness \textit{Recognizes} the Collapse Hypothesis (CRCH) \citep{LonBau1939theory}. So we have ontological underdetermination right from the start \textit{within} CBI. Let us focus on the CCCH only.

As famously argued by \citet[p.~421]{vNeum1955mathematical}, superposed states as in equation \ref{superpositionM-Z} collapse in an eigenstate of the observable through the interaction with the ``abstract ego'' of an observer; \citet{wigner1961mindbody} goes even further stating that the causal agent is human consciousness. The detection in $D_1$ or $D_2$ would not be sufficient to cause the collapse, because linearity implies that the detectors' states are to be described as superpositions as well. This situation occurs because the whole system described by $|\psi\rangle$ is in the same ontological domain: the \textit{physical domain}; so, in order to account for measurement outcomes, it can be only through the interaction of a \textit{non-physical} agent that breaks the superpositions and collapse the superposed state into a single-term state \citep[p.~421]{vNeum1955mathematical}, which is what we actually observe. This provides clearly for an ontology, although, as we mentioned, not for a clear metaphysical picture of what such consciousness is.

Now let us pass to QM$_{bra}$. It says that in the case described by equation \ref{superpositionM-Z}, the superposition is taken literally: both terms are \textit{equally real}. So, if we find, say, $|\psi_1\rangle$ as a measurement outcome, then the outcome $|\psi_2\rangle$ is also the case elsewhere.

We say \textit{elsewhere}, because QM$_{bra}$ does not determine its ontology: on the many-worlds interpretation (MWI), the branching splits the \textit{worlds} \citep{dewitt1971many}; on the many-minds interpretation (MMI), the branching splits the \textit{minds of the observers} \citep{Loc1989mind}. We will stick with the former only, but, for the record, we also have an ontological underdetermination at this point. Let us focus on the MWI only.

An ontological aspect of QM$_{bra}$ is MWI, in which the world branches itself into each possible outcome of a superposition. If we find, say, $|\psi_1\rangle$ as a measurement outcome, then we happen to live in this branch of the world --- the outcome $|\psi_2\rangle$ is still true for an alternative (equally real) branch of the world, where such measurement outcome is simultaneously found. So although the collapse is apparently happening (insofar as we don't actually get to \textit{see} a superposition), it is not \textit{really} going on --- what in fact happened is the branching of the world in two, and we do not perceive other worlds. And that's also part of the head of the dragon: we perceive determinate outcomes of measurement.

Given that quantum mechanics seems to lead us directly to ontological underdetermination, what are the prospects for any attempt to engage ontology and quantum mechanics in order to provide for a clear worldview?

\section{Naturalized Ontology} \label{sec:3}

Before we address the question that was left hanging in the air in the end of Sec.~\ref{sec:2}, a distinction between metaphysics and ontology is in order.

To many authors, contemporary analytic metaphysics is worthless as a guide to the objective features of the world \textit{unless} it can be continuously guided by contemporary science \citep{Ladyman2007evmustgo}. Such project is known as the \textit{radical naturalization} of metaphysics, which advances a skeptical attitude towards most of what currently falls under the label of ``analytic metaphysics''. This kind of debate suffers from some ambiguities on its scope and aims, a difficulty which could be overcome, we believe, if a proper distinction between ``ontology'' and ``metaphysics'' is advanced. Such a careful distinction could avoid many problems we find in current literature on the naturalization project.

One of the most common of such problems is the naturalistic expectation that scientific theories provide everything we need to know in terms of metaphysics and ontology, so that we can \textit{read off} or \textit{extract} a metaphysics (again, a term often treated as interchangeable with ``ontology'') directly from a given theory's scientific basis (see \cite{jonas2019filomena, RaoJonas2019dualismQM} for further discussions on this topic).

If for a moment we dissociate the two terms, we might be able to see the achievements and limits of the radical naturalization project (and this could work to the benefit of such project). As we will discuss in this section, naturalists may have reasons to invite ontology to their parties, but the same is not so clear for metaphysics.

In order to make this clearer, we use the results of Sec.~\ref{sec:2}. Take for example QM$_{col}$ and QM$_{bra}$. Both interpretations of QM postulate different entities as populating the world, which have a fundamental (but also different) theoretical role, but which do not appear directly in the mathematical formalism.

So, where are the entities in the ontology of such theories? Certainly not in the formalism, but surely within the theories' scope: in CCCH, for example, the causal agent of the collapse is human consciousness \citep{wigner1961mindbody}; in MWI, the branching process depends on the positing of multiverses. These entities are part of what we have been calling the \textit{ontology} of the theory, the entities that populate the world according to the theory. In this sense, we define the term ``ontology'' as that part of philosophy concerned with providing a \textit{catalogue} or \textit{list} for \textit{what there is} in the world (according to a given theory or interpretation of a theory). Entities such as human consciousness and multiverses \textit{should} be regarded as fundamental components of the ontology of CCCH and MWI respectively, just as particles such as protons and electrons, as observed by \citet{ladyman2019qontology}.

As \citet[p.~3433]{ruetsche2015shaky} pointed out, the interpretations provide precisely for the realist content of the theory; a realist who believes in QM believes in the posits afforded by an \textit{interpretation} of QM. On the one hand, a theorist inclined to the CCCH interprets the measuring act formally as collapse, and accounts \textit{ontologically} for the collapse by the introduction of a non-physical human consciousness with a causal power in measurement processes; on the other hand, a theorist inclined to the MWI interprets the measuring act as the branching process which \textit{ontologically} leads to physically distinguishable worlds after the measurement of every superposition state. These ontological claims state something concerning the very existence of the processes and entities that describe physical phenomena.

One meta-ontological method of extracting the ontology from theories could be a Quinean-like approach of \textit{ontological commitment}. Recall that, to \citet[p.~65]{quine1951commitment}, this means simply ``\textelp{} what, according to that theory, there is''. So, despite Quine's own eventual conflations between ontology and metaphysics (see \cite{quine1969existencequantif}), his discussions of ontology and metaphysics always seem to concern ``ontological commitment'', i.e., questions of existence. So, when we look at the entities that play some fundamental theoretical role in scientific theories, we may say that those theories are committed with the existence of such entities. Therefore, even though the term ``consciousness'' does not have any mathematical counterpart in the formalism of CCCH, it plays a fundamental theoretical role as the causal agent of the collapse; so it is safe to say that CCCH is \textit{ontologically committed} with the existence of consciousness, and hence it should be regarded as part of the theory's ontological catalogue; the same goes for the many worlds in MWI.

Other approaches to ontological commitment may be available. Our main claim is not that Quinean approach to ontology is to be adopted, but to illustrate that it concerns exclusively existence questions. Furthermore, our point is also that once one adopts an interpretation of quantum mechanics such as those discussed previously, it would be idle to reject commitment to consciousness and to many-worlds, respectively. Those items are present in the furniture of the world of those adopting such interpretations.

The idea that the ontology of quantum mechanics is associated with its interpretations may be further illustrated in another Quinean move. There is a view which consider that it is possible to extract at least some fundamental ontological aspects from the formalism alone. This view is called ``wave function realism''. Such view is sometimes presented as a natural attitude towards QM \citep{albert2013wavefunction}. Typically, it is based on the famous Quine-Putnam argument for the indispensability of mathematical entities \citep{colyvanSEPindispens}. The indispensability argument is generalized for obtaining an ontology from the very formalism of QM --- that is, the ontology can be ``read off'' directly from the formalism. A scheme of the argument was advanced by \citet[p.~67]{ney2012neopositivist} as follows.

\begin{itemize}
\item[$P_1$] We ought to have ontological commitment to the entities that are indispensable to our best scientific theories.
\item[$P_2$] Entities of kind X are indispensable to QM.
\item[$\therefore$] We ought to have ontological commitment to entities of kind X.
\end{itemize}

As the wave function ($|\psi\rangle$) representation is indispensable to the quantum-mechanical description of quantum systems, we then ought to have ontological commitment to the wave function \textit{as an entity} in the theory's ontology. For the sake of the example, take the \textit{which-path} description of the Mach-Zehnder experiment, as shown above in equation \ref{superpositionM-Z}. It describes the motion of the wave function $|\psi\rangle$ through time, so one should include the entity ``wave function'' as an existing entity \textit{modulo} QM. Whether the wave function realism should be regarded either as an interpretation of QM, in addition of QM$_{col}$ and QM$_{bra}$ (see \cite{albert2013wavefunction}), or as a general attitude towards the ontology of QM which is independent of the theoretical choices one might make to solve the measurement problem (see \cite{plewis2004configuration, ney2013introduction}) is a matter that we shall not discuss here, for our point does not depend on taking a stance on such issues. At least from the point of view of wave function realists, \textit{some} part of the ontological catalogue is \textit{read off} directly from the theory's formalism.

Once this is in order, a question naturally emerges. We have spoken of ontological underdetermination in Sec.~\ref{sec:2}, and clearly there are many incompatible interpretations of quantum mechanics, all of them populating the world with distinct processes and entities. That is, the questions of existence, of ontology, receive not a single answer from quantum mechanics, given that their answer is related to an interpretation, and interpretations are many. How is it possible to talk about naturalization of ontology, in the face of such an underdetermination? Well, the naturalist has a way out, we think. Let us briefly discuss it (this is really a very brief treatment of the subject, which we shall address with more details in a future work). The main challenge consists in choosing one of the rival interpretations. But, the naturalist may argue, this is a scientific matter, one that scientists face every now and then. Quantum mechanics is still an ongoing research project, and it is no wonder that some options concerning which version is the correct one are still on the table. That happens naturally in scientific research. Sooner or later some of them will be eliminated as non-options, and others will remain. If there is not a single interpretation adopted, that is still not a problem. Furthermore, and even more to our point, the naturalist may also advance that such refining of the options, advancing interpretations, in the case of quantum mechanics, is a matter of doing more physics. As  \citet[pp.~8--9]{Maudlin1995measurementproblem} has argued, advancing one interpretation as the correct one (and with that, advancing an ontology as the correct one) requires doing more physics, this is not an issue left for sheer philosophical speculation. In this sense, then, a naturalization of ontology has some good perspectives, given its close ties to the development of physics itself.

What is relevant for us is that given an interpretation, it is more or less clear what kinds of entities it does posit. Ontology may be obtained from the interpretation, although there may be space for controversy over some of the details. If we cannot have scientific reasons for choosing one ontology over the other, that is not for philosophical reasons, but due to science itself, at its current stage. Given that naturalists cannot get out of science to make philosophy, the best they can do is to investigate the available scientific options.

So that's it, right? The theory gave us its ontology, so that we have our science-based worldview obtained in a completely naturalistic fashion: \textit{science tells us what the world is like, and according to QM, the world is such and such.} Is the task of the metaphysician over? Well, not quite so, according to many metaphysicians. In the sense of `ontology' defined earlier, we may have extracted our ontologies from the distinct interpretations of quantum mechanics --- either directly from formalism or otherwise --- which is an account of what there is \textit{modulo} such theory. But a further question remains: what are such things that populate the catalogue of reality? In other words, the main point can be summed up in the following question: is the ontology obtained also informative about the metaphysical profile attributable to those entities? And the answer is straightforward: not at all. Let us advance that point.

\section{Enter the Viking Approach} \label{sec:4}

The sense that something is still missing on a clear picture of reality may have assaulted some readers of Sec.~\ref{sec:3}. Ontology --- as previously defined --- is not enough for an appropriate account of reality. Regarding wave function realism, \citet[p.~68]{ney2012neopositivist} states that \textit{even if} we conclude that the wave functions exist, we ``\textelp{} do not have justification for making factual assertions about what sort of entity this wave function is''. We claim that the same holds precisely for the consciousness-based or world-splitting accounts of the universe\footnote{And also of posits of other interpretations not being discussed here, of course.}: nothing is said about the sort of entities consciousness or splitting worlds are. So QM (QM$_{col}$, QM$_{bra}$ or else) does not tell us what are, in metaphysical terms, the entities with which they are ontologically committed. Nevertheless, our claim is somewhat distinct from Ney's perspective. According to her, ontology is literally \textit{read off} from the theory, and the ``metaphysical extras'' are part of the theory, although added `by hand', as it were. On the other hand, we propose that such metaphysical characterization is an \textit{extra layer} over the theory itself. This shift on the perspective has some important methodological implications: while on Ney's perspective the metaphysical description of the entities is part of the scientific theory itself, although a part that does not derive any direct justification from the theory, on our view the metaphysical layer is proposed by philosophers of science after the development of scientific theory; metaphysics, as an extra layer, is not used \textit{directly} in scientific theory, so metaphysicians are not the bearers of novelties that can be decisive for scientific development; the metaphysical layer consists only of an additional explanation, which is constructed from the scientific and ontological explanation.

We shall discuss this in a moment. For now, what matters is that the lack of a metaphysical characterization \textit{derived} from the ontology might be a drawback for those endeavouring to look at QM when it comes to tell us what the world is like --- which some say is a cornerstone of scientific realism. To \citet[p.~48]{french2014structure}, the standard approach to scientific realism is summed up in the following ``recipe'': ``\textelp{} we choose our best theories; we read off the
relevant features of those theories; and then we assert that an appropriate relationship holds between those features and the world''.

Such a recipe, however, seems to leave out important aspects for a finer understanding of the world --- including a \textit{metaphysical} image of what the world is like \textit{modulo} any given scientific theory. Here, by ``metaphysics'' we understand the philosophical discipline concerned with the most general structure of reality, so that once the posits of a scientific theory are identified (in the corresponding ontology, as explained in Sec.~\ref{sec:3}), metaphysics still has a job to do by providing them what may be called a ``metaphysical profile'', by describing what they are in metaphysical terms. As we shall discuss soon, metaphysics, as we have already mentioned, adds an extra layer of theory over the posits of ontology.

\citet{french2013handling} is perhaps one of the most well-known proponents of the very idea that a metaphysical profile must be provided on the top of the posits of a theory (i.e, the theory's ontology), on the pain of lacking a clear image of the posits of scientific theories. His own approach is presented by a threefold requirement:

\begin{enumerate}
    \item[(i)] The metaphysical gap must be filled;
    \item[(ii)] It can be filled with an outlook in the history of philosophy;
    \item[(iii)] Such gap-filling enables us to understand the world better than we did without it, so we may reduce our epistemic humility.
\end{enumerate}

In what follows, we will assume those three aspects of French's view on scientific realism and metaphysics as a working hypothesis and see how well they fare in giving us a proper methodology for understanding the relation between metaphysics and science. This, as we have already commented and as the reader may have noticed, involves adding an extra metaphysical layer over the ontological posits of a scientific theory. So, we build from there.

Let us begin with the first aspect: that \textit{the metaphysical gap must be filled}. Where did the need for a metaphysical dressing of the entities come from? \citet[p.~26]{chakravartty2007metaphysics} wrote that: ``One cannot fully appreciate what it might mean to be a realist until one has a clear picture of what one is being invited to be a realist about''. This claim, it is said, calls for the need of a metaphysical explanation, the so-called \textit{clear picture} that explains \textit{what are those entities}, and this explanation must be provided for in metaphysical terms (or so it is said).

This further sense of interpretation concerns what we are calling here the attribution of a ``metaphysical profile'' (see also \cite{jonas2019filomena}). But it should be clear that obtaining such clear picture of the metaphysical profile is not so simple: it should be handled, as \citet{french2013handling} pointed out, with some degree of epistemic humility (as we do not have something like a point of view of nature from a \textit{cosmic exile} to evaluate between rival options in metaphysics). Handling the epistemic humility with the need of a metaphysical profile (the ``clear picture'') was called by \citet[p.~85]{french2013handling} ``Chakravartty's Challenge''. According to the challenge, it is not enough to point to some feature of a theory and claim realism about it.

In order to have a legitimate realism about the contents of a scientific theory, one must clearly specify what it is, and doing so involves --- at least partially --- providing for a metaphysical characterization of such content. So, suppose that one claims to be realist about, say, CCCH. This means, among other things, that such person believes that CCCH makes true statements about the world in which we live in, \textit{including} claims about its unobservable processes --- such as the causal power of consciousness (with which the theory is ontologically committed). However, if no metaphysical profile is offered for consciousness, the alleged realism about the theory is empty (or so the kind of realist we are examining now says).

\citet[p.~8]{bueno2019viking}, for instance, resists Chakravartty's Challenge by resisting ``\textelp{} the temptation to reify what is posited in one's ontology''. Once that is denied, he feels free to pursue an empiricist approach to quantum mechanics. And that's precisely the most radical implication of Chakravartty's Challenge that we are stressing here: unless one is willing to become an empiricist, it seems, the search for a metaphysical profile is a non-trivial requirement --- at least if the debate is accepted in its current terms (and again, we are accepting these terms for the sake of argument).

With this in mind, one may be naturally led to the question: how to obtain a metaphysical profile (\textit{e.g.}, Chakravartty's ``clear picture'') for a theory's ontological content? This leads us to the second aspect of the hypothesis made by \citet[chap.~3]{french2014structure}: \textit{that the metaphysical gap can be filled with an outlook in the history of philosophy}.

Can philosophers just come up with metaphysical characterizations for the entities that scientific theories are ontologically committed to? Well, sure. But \textit{are we justified} to ask for metaphysicians to do such a job? The question is sensible because, as \citet{french2011metaphysical} worries, if such metaphysical theory ``floats free'' from scientific theories, then metaphysics won't find any justification in science. Moreover, such worry implies that we might run the risk of driving metaphysical inquiries which are totally distant from the empirical results, and thus these inquiries would themselves require justification. This situation of total lack of relation to science has led many authors, such as \citet{Ladyman2007evmustgo}, to argue for a discontinuation of analytic metaphysics. Worse yet: without such an anchorage in science, one runs the risk of metaphysical underdetermination. But let us proceed one step at each time.

First of all: can philosophy fill the metaphysical gap? Here is the catch: Yes; but metaphysics, in the context of the requirement of a metaphysical profile, must float free from physics. It must come from somewhere else than science itself. Otherwise, it would be impossible to \textit{attribute} a metaphysical profile to QM --- and, according to Chakravartty's Challenge, it would be impossible to claim realism about QM. French himself acknowledges that, when stating that

\begin{quote}
\textelp{} you get only as much metaphysics out of a physical theory as you put in and pulling metaphysical rabbits out of physical hats does indeed involve a certain amount of philosophical sleight of hand. \citep[p.~466]{french1995rabbit}
\end{quote}

Indeed, if metaphysics were not an additional layer cast over the posits of the theory, and could be somehow found out as somehow already hidden in the theory itself, then there would be no need for a philosophical investigation of it. It would be just another physical fact. But, as we know, it is  not.

Being part of the physics, it would make no sense to look for metaphysical concepts to dress the posits, the problem would be ill-conceived, and this seems to be the epistemic dilemma for the scientific metaphysicians. So the quest for realism, through Chakravartty's Challenge, justifies the use of metaphysical tools which are not necessarily scientifically informed --- the so-called \textit{armchair methods} \citep[p.~10]{Ladyman2007evmustgo}.

In such project, there are authors who acknowledge the utility of metaphysics to scientific theories precisely in filling such metaphysical gap, being, in principle, useful for the interpretative work: so, analytic metaphysics may be regarded as a ``toolbox'' \citep{french2012toolbox} that is used to interpret scientific theories, or a ``spot for pillaging'' \citep[p.~v]{french2014structure}. The last suggestion is the ``Viking Approach'' to metaphysics, which justifies the free development of metaphysics based on the scientific usage of metaphysical concepts. To the Viking Approach, in French's own words, ``\textelp{} the products of analytic metaphysics can be regarded as available for plundering!'' \citep[p.~50]{french2014structure}.

Chakravartty's Challenge is thus addressed by the Viking Approach in the following way: one simply chooses among the available options in the metaphysics' literature in order to attribute to the posited entities a metaphysical profile. In this sense, metaphysics, as a discipline, has roughly the same level of independence from empirical science as mathematics, for it provides theories, tools and strategies of investigation and speculation that can be used in order to interpret scientific theories. A similar argument for the independence of \textit{a priori} metaphysics from empirical sciences, and the analogy with pure mathematics, can be found in \citet{morgantitahko2017moderately}. An interpretation of a scientific theory, such as QM, then, not only gives the realist content of the theory (see again \cite{ruetsche2015shaky}), but, following Chakravartty's Challenge, opens up the \textit{possibility} of obtaining a realist content through the use of a metaphysical profile.

However, at least one question still remains unanswered: when is the challenge successfully met? Recall that it is not possible to extract metaphysical profiles from scientific theories \textit{and} it is not possible to be realist about something without its metaphysical profile (or so the challenge goes). But, as \citet[p.~12]{chakravartty2019challenge} recalls, ``[i]t is \textit{always} possible to ask finer-grained questions \textelp{}''. So when is the metaphysical profile sufficiently developed in order to justify realism about the theory it sits on the top of? Let's call this the ``Meta-Chakravartty's Challenge''. Nevertheless, by recalling that the \textit{clear picture} that the challenge demands can only be obtainable with a \textit{floating-free} metaphysics, the Meta-Chakravartty's Challenge shows itself to be not so shocking, and becomes more naturally acceptable (at least for those already accepting the kind of scenario we are discussing here): after all, it is indeed always possible to investigate finer-grained metaphysical questions --- this is most of the metaphysical work as well as its prerogative.

\citet[p.~394]{French2018RealMetaph} also acknowledged this problem, by asking ``how much further should the realist go'', and ``to what extend should our realism be metaphysically informed?''. The problem then returns to the question of epistemic humility \citep{french2013handling}: as metaphysical claims cannot be directly backed by observation (\textit{e.g.}, no physical theory can tell us, say, whether the concept of ``mass'' is an instantiated \textit{universal} or a \textit{trope}), so we must take a humble attitude towards metaphysical profiles. How much humility one should adopt, however, is unclear: if too much, there is the risk of telling the same story that relevant scientific theories already does (thus not forming a worldview); if too little, then one may have to face underdetermination (many metaphysical options are available, with no physical prospects of choosing one of them in physically acceptable terms).

This leads us to the third aspect on French's view: epistemic humility. To \citet[p.~401]{French2018RealMetaph}, the perfect balance for epistemic humility lies in his Viking Approach, which is close enough to the ``Toolbox'' \citep{french2012toolbox} approach: the metaphysical literature provides us with many points of view, strategies, and approaches that some may use in articulating a metaphysical profile to the relevant characteristics of the theory in which they believe. According to this view, we may think of the CBI within a Cartesian-like dualism or a Husserlian-like phenomenology. So, Chakravartty's Challenge (and the Meta-Chakravartty's Challenge also) are completed with the \textit{use} of metaphysical theories, developed by metaphysicians, in a way that the realist can employ the tools in articulating the questions concerning \textit{how the world is} --- \textit{modulo} CCCH or some other interpretation of QM.

Of course, this is an heuristic move, because one may articulate, by itself, such a clear picture. But then, again, \citet[p.~404]{French2018RealMetaph} asks, ``why reinvent the wheel?'' --- to what we fully agree: it is more fruitful to first take a look on the available options. So, French's method of approaching metaphysics within the context of science, is not an anecdote; the Viking and the Toolbox approaches are not just ways of saying, with a dismissive tone, something like `let's look at a lot of things that metaphysicians have done', but is rather a literal suggestion for taking the realist requirement of a clear picture seriously. About such suggestion, French states that the way of attributing a metaphysical profile to scientific concepts is:

\begin{quote}
\textelp{} to engage with extant metaphysics, draw on the tools it has already developed, and work with metaphysicians themselves to hone and sharpen them in various ways, so that they can be developed more precisely to help us understand what it is that science is telling us. \citep[p.~405]{French2018RealMetaph}
\end{quote}

And, looking at metaphysics of science in this way, it does not seem so far from traditional, armchair metaphysics. A first and somewhat obvious reason is that, if the Viking Approach to metaphysics uses what metaphysicians already have done in the history of philosophy, then much of what is plundered are products of this traditional metaphysics (in which, for example, Plato's theory of forms is still an available source for pillage). A second, and perhaps speculative reason for such conclusion is that the so-called metaphysics of science does not seem to have produced metaphysical content, and remains thus far as a methodological attitude \textit{towards} metaphysics --- a reason why it might be more prudent to call it a \textit{metametaphysical} attitude.

\subsection{Evaluating the metaphysics of science}

Consider for a moment French's three-folded strategy (Chakravartty's Challenge, the Viking Approach and the decrease of Humility). By calling for the need of a metaphysical profile, the best we can get with our epistemic limitations is to \textit{plug} some already-existent metaphysical theory to the entities that are part of the \textit{ontological catalogue} of a given scientific theory.

The first bad news is that humility is not really decreased. If metaphysics is not tied up directly with science, \textit{i.e.}, if it \textit{floats free}, then metaphysical underdetermination must be taken for granted. After all, it is conceivable --- indeed, it is very likely --- that more than one metaphysical theory can serve to provide for a metaphysical dressing for a single entity. This is the main virtue of the Viking approach, and also, its main vice.

This `floating free' aspect of metaphysics, however, by no means discards the influence of the empirical results. It is worthwhile to mention that in recent paper, \citet[p.~28]{FRENCH2019defending} recognized this issue, when he states that, if metaphysics wants to say something relevant about the world in which we live in, then ``\textelp{} appropriate consideration needs to be taken of the relevant physics and the constraints it imposes. What follows is a short account of the argument more fully discussed in \citet{RaoJonas2019dualismQM}. For the sake of an example, take CCCH: it states that consciousness \textit{causes} a modification in the dynamics of quantum systems, so it constrains the available metaphysical profiles for such entity, consciousness (see also \cite{jonas2019filomena} for further discussion). Whatever it may be in metaphysical terms, ontology requires that consciousness must be real, with causal powers. Several metaphysical profiles in which it has no such feature are ruled out: they are \textit{incompatible} with CCCH's ontological catalogue. A fairly obvious example is epiphenomenalist metaphysical profiles, which clearly fails to cope with CCCH's ontological requirements for metaphysics: the features of \textit{reality} and causality of consciousness.

\section{Final remarks} \label{sec:5}

Summing up, from our discussions we may draw two important conclusions for the methodology of the metaphysics of science. First of all, those accepting that the problem of providing a metaphysical profile, as advanced by Chakravartty's Challenge, poses a legitimate question, will have to concede that the metaphysics floats free from physics; second, once it is assumed the kind of scenario in which Chakravartty's challenge is framed, the fact that a metaphysical profile is provided, taken by itself, does not reduce humility. This happens because, since metaphysics is not supported by science, metaphysical underdetermination must be taken for granted from the outset. This puts some pressure on the very idea that the metaphysical profile could derive some justification by its association with science. Mere association is not enough, in the face of the fact that the metaphysics really is floating free from the science.

Taking into account what has been said so far, we propose below some points for reflection.

Weighing pros and cons, a Viking approach to the metaphysics of science has a very clear gain: it recognizes, and somehow requires, that metaphysics has an autonomy as a philosophical discipline, in relation to scientific disciplines. The price of this is that it stands in a disturbing distance from science; that is, it doesn't take any part in science's epistemic credentials. Metaphysics may provide a worldview, but metaphysical theory choice cannot benefit from science. Since metaphysics floats free, the image of metaphysics according to the Viking approach ends up somehow picturing a disengaged metaphysics. Regarding this methodology, the idea of a discontinuation, or at least (and, perhaps, more productively) a reformulation of metaphysics in relation to science, is pertinent. That is, for those expecting a productive relation between science and metaphysics, a kind of scientific metaphysics --- in opposition to the metaphysics of science --- must be sought.

\section*{Acknowledgments}
An earlier version of this paper was presented as a talk at the 11th Principia International Symposium (\url{http://www.principia.ufsc.br/SIP11.html}), held in Florian\'opolis, Brazil, in the year of 2019. We would like to thank the event attendants, whose constructive suggestions contributed to the improvement of this text. We would also like to thank Christian de Ronde for inviting us to participate in this volume, and for the always thought-provoking conversations about the philosophy of physics and scientific realism. Finally, we would like to thank Steven French for the joy of discussing his great work, which keeps constantly inspiring philosophers and metaphysicians of science.

\printbibliography

\end{document}